\documentclass[11pt]{article}

\usepackage[left=1in,right=1in, top=1in, bottom=1in]{geometry}

\usepackage{amsthm}
\usepackage{amsmath, amssymb}

\newtheorem{fact}{Fact}

\newtheorem{theorem}{Theorem}
\newtheorem{corollary}{Corollary}

\usepackage[numbers,sort&compress,longnamesfirst,sectionbib]{natbib}             

\newcommand{\e}{\varepsilon}
\newcommand{\eps}{\varepsilon}

\DeclareMathOperator{\diam}{diam}

\newcommand{\N}{{\mathbb{N}}}

\usepackage[ruled, vlined]{algorithm2e}
\usepackage[bottom]{footmisc}

\title{Asymptotically Optimal Randomized Rumor Spreading}
\author{Benjamin Doerr\footnote{Department 1: Algorithms and Complexity, Max-Planck-Institut f\"ur Informatik, Saarbr\"ucken, Germany} \and Mahmoud Fouz\footnote{Faculty of Computer Science, Universit\"at des Saarlandes, Saarbr\"ucken, Germany}}

\begin{document}

\maketitle


\begin{abstract}
  We propose a new protocol solving the fundamental problem of disseminating a piece of information to all members of a group of $n$ players. It builds upon the classical randomized rumor spreading protocol and several extensions. The main achievements are the following: 

Our protocol spreads the rumor to all other nodes in the asymptotically optimal time of $(1 + o(1)) \log_2 n$. The whole process can be implemented in a way such that only $O(n f(n))$ calls are made, where $f(n)= \omega(1)$ can be arbitrary. 


In contrast to other protocols suggested in the literature, our algorithm only uses push operations,
i.e., only informed nodes take active actions in the network. To the best of our knowledge, this is the first randomized push algorithm that achieves an asymptotically optimal running time.
\end{abstract}

\thispagestyle{empty}
\pagebreak
\setcounter{page}{1}
\section{Introduction}
Transmitting a piece of information to all nodes of a network is a classical problem in computer science. A protocol surprisingly powerful is called \emph{randomized rumor spreading}, see, e.g.,~\citet{FG85,FPRU90,KSSV00}. It proceeds in rounds as follows: in each round, each node that already knows the piece of information (``rumor'') chooses a communication partner uniformly at random and sends her a copy of this rumor.  

In spite of being that simple, this protocol succeeds in spreading the rumor to all nodes of a complete graph in $(1 + o(1))(\log_2 n + \ln n)$ rounds with high probability, that is, with probability $1 - o(1)$. In addition, due to its randomized nature, it is highly robust against different types of transmission or node failures. This makes it an interesting alternative to deterministic protocols, which can reduce the broadcast time to $\log_2(n)$, but at the price of suffering greatly from failures.

A clear disadvantage of this most simple version of randomized rumor spreading is the enormous number of $\Theta(n \log n)$
calls that are necessary. This problem was overcome in the seminal work of \citet{KSSV00}. They present two variations of the randomized rumor spreading protocol which spread the rumor with a total number of $O(n \log\log n)$ messages only while still using $O(\log n)$ rounds only. A central ingredient are so-called pull operations, which allow nodes not yet informed to call random nodes and ask for news. Pull operations, however, have the disadvantage that they create network traffic even if there is no news to be spread. Therefore the assumption underlying the analysis of \citet{KSSV00} is that there is constantly new information injected in the network.

In this work, we present an alternative solution to the problem. It completely avoids the problematic pull operations. It achieves
a broadcast time of
$(1+o(1)) \log_2 n$ and it uses a total number $O(n f(n))$ calls, where $f = \omega(1)$ can be 
any function tending to infinity arbitrarily slow. This is very close to the theoretically optimal values of $\lceil \log_2 n\rceil$ rounds and $n-1$ calls. 
Due to its randomized nature, we still have reasonable robustness. For
example, if a constant fraction of the nodes chosen uniformly at random crashes at arbitrary times, the time needed to inform all properly working nodes
increases by at most a constant factor (depending on the failure rate). 

The only point in which we assume the protocol to be more powerful compared
to previous works
is that we discard the address-obliviousness. That is, we assume that each node has a unique label chosen arbitrarily from some
ordered set (e.g., the
integers). This seems to be a reasonable assumption in many settings.

\subsection{The protocol of \citet{KSSV00}}

As described above, \citet{KSSV00} showed how to modify the simple randomized rumor spreading protocol such that instead of $\Theta(n \log n)$ messages only $O(n \log\log n)$  are sufficient to spread a rumor. Roughly speaking, their protocol proceeds as follows. The rumor is equipped with a time stamp (or age counter) in such a way that all nodes that receive the rumor also know for how many rounds it has been in the network. In each round, each node chooses a random other node as a communication partner. The communication then proceeds in both directions, that is, any partner who knows the rumor forwards it to the other partner. It is shown that after $\log_3 n + \Theta(\log\log n)$ rounds of this protocol, all nodes know the rumor with high probability. In addition, a rumor is transmitted in this time interval at most $O(n \log\log n)$ times. 

Note that this way of counting tacitly ignores all communication effort which does not result in a rumor to be sent. In particular, all calls between two uninformed nodes that arise due to pull operations are ignored. The way this
is usually justified is by assuming that there is sufficient traffic in the network due to regular insertions of new rumors. Still, we feel that this
is slightly dissatisfying. Note that when using pull operations, there is no way to avoid such communication overhead---a node that did not
receive a rumor recently has no way of finding out whether there are rumors around that justify starting pull operations or not. Moreover, even nodes that did receive a rumor recently cannot be sure that there is no new rumor that would justify starting pull operations again. 


\citet{KSSV00} also prove lower bounds, which, roughly speaking, show that if in each round all communication is restricted to random matchings of communication partners (i) any address-oblivious algorithm has to make $\Omega(n \log\log n)$ calls and (ii) that any algorithm informing all but a $o(1)$ fraction of the vertices in logarithmic time has to make $\omega(n)$ calls.

\subsection{Our results}

The first lower bound stated in the previous paragraph suggests that asking for an address-oblivious protocol may result in only a
limited performance being achievable. In addition, one might also wonder if really many 
broadcasting problems ask for
address-oblivious protocols, or if not rather in the majority of settings each participant naturally has a unique addresses, simply to
organize the transport of a message to an addressee.

In this work, we shall skip the requirement of address-obliviousness. However, we shall keep the concept of contacting random neighbors without any preference, as this seems to be the key to obtaining good broadcasting times, robustness and small number of 
calls in all previous works.

Contrary to the model of \citet{KSSV00}, we do not perform pull operations. That is, all transmissions are initiated by nodes that
know the rumor. In consequence, the only direction of informing is from the initiator of the transmission to its addressee, which is chosen uniformly at random, though not always independently. 

We do allow, however, two-way communication, in that the addressee acknowledges his readiness to receive the rumor or the fact that he already knows the rumor. Such a mechanism makes sense anyway, because it allows to reduce the amount of data sent through the network (if the addressee cannot receive the rumor or already knows it, we do not need to send it). In practice, most communication protocols (e.g., the standard network protocol FTP) allow some kind of two-way communication to ensure an error-free transmission.

In this, as we think, natural setting, we propose a protocol that needs only $(1 + o(1)) \log_2 n$ rounds and $n f(n)$ calls, where $f = \omega(1)$ can be chosen arbitrarily. Note that no protocol that only uses push operations can work in less than $\lceil\log_2 n\rceil$ rounds or using less than $n-1$ calls.

More precisely, we have the following trade-off between rounds and messages. For all $f: \N \to \N$, we give a protocol that needs only
$\log_2(n) + f(n) + O(f(n)^{-1} \log n)$ rounds with high probability and $O(nf(n))$ calls. In terms of run-time, this is optimal for $f(n) = \Theta(\sqrt{\log n})$, leading to $\log_2 n + O(\sqrt{\log n})$ rounds and $O(n \sqrt{\log n})$ calls. 

The protocol is very simple. For the presentation, let us assume that the nodes are numbered from $1$ to $n$,
even though what we really need is only that nodes are able (i) to compute the label of a node chosen uniformly at random and (ii) given a label of a node, to compute a uniquely defined successor along a cyclic order of the labels (here the label plus one, modulo $n$). 

Let $f: \N \to \N$ be given (to formulate the tradeoff scenario). Then the protocol works as follows. Each newly informed node sends its first message to another node chosen uniformly at random. From then on, it does the following. If the previous message was sent to a node that was not informed yet, then the next message is sent to the successor of that node in the cyclic order. Otherwise, the next message is sent again to a node chosen uniformly at random. After having encountered $f(n)$ nodes that were already informed,
the node stops and does not transmit the rumor anymore. This protocol can be interpreted as a variant of the quasirandom rumor spreading protocol investigated in~\cite{DFS08,DFS09}, where in addition all nodes have the same cyclic permutation and they re-start at a random position whenever they call a node that is already informed (up to $f(n)-1$ times).

The main technical difficulty in the analysis of the proposed protocols stems from the fact that the transmission of messages at each node is not independent, and thus, many classical tools cannot be employed. The key to the solution here is to exploit the existing independence stemming from communications started with random partners.

In summary, our result shows that considerable improvements over the fully independent rumor spreading protocol are possible if we skip the requirement that the protocol is address-oblivious. It thus seems worthwhile questioning whether the address-obliviousness assumption is really needed in previous applications of the protocol. From the methodological side, our result again shows that spicing up randomized algorithms with well-chosen dependencies can yield additional gains. It may make the theoretical analysis more complicated, but not so much the algorithm itself. 

\subsection{Disclaimers}

\textbf{Applications:}
For reasons of space, we have not given extensive details on randomized rumor spreading and its applications. The seminal papers
\citet{FPRU90} and
\citet{KSSV00} contain great discussions of this, better than we could possibly do here. For reports on the actual use of such
protocols, see
\citet{Hedetniemi88,DGH+87} and \citet{KDG03}.\\\newline
\textbf{Other network topologies:}
Randomized rumor spreading can be used on all types of network topologies. Nodes then choose their communication partners at random from the set of their neighbors.
For many network topologies, broadcast times logarithmic in the number of nodes have been shown. Besides the complete graph, they
include hypercubes~\cite{FPRU90}, random graphs $G(n,p)$ with $p \ge (1 + \eps) \ln(n)/n$~\cite{FPRU90} and certain expander
graphs~\cite{S07}. Recently, rumor spreading was also shown to be doable in poly-logarithmic time for social networks modelled by preferential attachment graphs~\cite{CLP09} and for graphs of bounded conductance~\cite{CLP10}. For Cayley graphs~\cite{ES07} and random geometric graphs~\cite{BEFSS10}, the (in another sense) near-optimal bounds of $O(\diam(G) + \log n)$ are known. In spite of these results, we concentrated ourselves on the setting where each node has a direct way to communicate with each other node. The reason is that we feel that this is a sufficiently interesting and useful case on its own. Also, of course, it is the setting in which it is
easiest to experiment with new ideas. Recall that the concept of reducing the number of messages was also first demonstrated on complete graphs by~\citet{KSSV00}. Only much later, similar results were obtained for other network topologies, e.g., by \citet{BEF08,E06} and \citet{ES08}.

\section{The Hybrid Protocol}

Let $G = (V,E)$ be the complete, undirected graph on $n$ nodes. We assume that the nodes
of the complete graph are ordered and denote by $i$ the $i$-th node according to that order. Our goal is to spread a `rumor' known initially to one node to all nodes in $V$. We call the node initiating the rumor the \emph{starting node}. A rumor can be transmitted along each edge of the graph in both directions. Every transmission along an edge is always initiated by a node that knows the rumor. We count every contact of a node to another node as a \emph{call}. We assume that two nodes never call a node exactly
simultaneously even if they both call the same node in the same round. Hence, a node is always only informed by a single node. 
   
We introduce a simple algorithm that for a certain instantiation achieves, up to lower order terms, an optimal running time. The algorithm is related to the quasi-random protocol by \citet{DFS08}. In this quasi-random protocol, every node $v$ is equipped with a cyclic permutation $\pi_v : V \to V$ of all nodes in $V$. Once a node $v$ becomes informed, it chooses one position on its list uniformly at random. This is the node $v$ contacts first in the next round. In each following round, $v$ contacts $\pi_v(u)$ where $u$ is the node it contacted in the previous round. Note that different nodes can have different permutations.     

Our hybrid protocol differs from this quasi-random protocol in two aspects. First, we assume that the permutations of all nodes are identical. Second, we introduce the notion of a \emph{restart}: if a node calls an already informed node, it chooses a \emph{random} communication partner in the next round instead of informing the next node according to the permutation. Furthermore, each node stops informing if it calls an informed node after the $R$-th random call, where $R$ can be a function of $n$. By this rule we can bound the total number of calls made. This aspect of keeping the number of calls small was not discussed in \citet{DFS08}. 

A detailed description of the hybrid protocol is given in Algorithm \ref{alg:basic}. The only exception is the starting node. This node does not select a random communication partner at the beginning, but starts informing its own successor node (according to the given permutation) immediately. Only after it encountered the first informed node does it then proceed according to Algorithm \ref{alg:basic}.   

\begin{algorithm}
\caption{Procedure started by newly informed node in the hybrid protocol}
let $R \in \mathbb{Z}^{+}$ be number of random calls per node\\ 
\For{$i = 1$ \emph{\KwTo}$R$}{select node $j$ uniformly at random\; \While{$j$ not informed \tcp{iteration counts as call even if $j$ informed}}{inform $j$\; $j \leftarrow j+1$\;} }
\label{alg:basic}
\end{algorithm}

We will analyze how long it takes for a rumor initiated by any node to reach all nodes under the proposed protocol. We give almost matching upper and lower bounds. Apart from the running time, we will also analyze the number of calls made.

\subsection{Running Time And Number of Calls}

We give an upper bound and an almost matching lower bound on the number of rounds and calls needed by the protocol to spread a rumor from an arbitrary starting node to all nodes of the complete graph.

\subsubsection{Upper Bound}

\begin{theorem}
\label{thm:tradeoff}
 Let $\e > 0$ be an arbitrarily small constant. With probability $1-o(1)$, the hybrid model with $R$ random calls per node informs all nodes in  
\begin{align*}
&\log_2 n + (1+\e) \ln (n)/R + R + h(n),& \text{if $R \leq \sqrt{\ln n}$}\\
&\log_2 n + (2+\e) \sqrt{\ln (n)},& \text{if $R \ge \sqrt{\ln n}$}
\end{align*} rounds, where $h(n)$ is a function of arbitrarily slow growth.  
This uses $n (R+1)$ calls.
\end{theorem}
Note that by adjusting the stopping parameter $R$, we get a tradeoff between the number of rounds needed to inform all nodes and the number of calls.


Before analyzing the protocol for general $R$, we describe two special cases that achieve an (almost) optimal number of rounds and
communications, respectively. 
For $R = \sqrt{\ln n}$, we achieve, up to a lower order term, an optimal running time while using only $O(n \sqrt{\ln n})$ calls.   
\begin{corollary} 
\label{upper_bound_time_efficient} 
Let $\e > 0$ be an arbitrarily small constant. With probability $1-o(1)$, the hybrid model with $R=\sqrt{\ln n}$ informs all nodes in $\log_2 n + (2+\e)\sqrt{\ln n}$ rounds. This uses $2(1+2\e)n\sqrt{\ln n}$ communications.
\end{corollary}
For $R = 1$, we get a very simple broadcasting protocol that, up to constant factors, is both
optimal in terms of rounds needed as well as the number of calls. 

\begin{corollary}
 \label{upper_bound_message_efficient} Let $\e > 0$ be an arbitrarily small constant. With probability $1-o(1)$, the hybrid model with $R = 1$ informs all nodes in
$\log_2 n + (1+\e)\ln n$ rounds. This uses $2n$ calls.
\end{corollary}

Before we prove Theorem \ref{thm:tradeoff}, we make two observations that will prove useful for the analysis.

\begin{fact}
\label{fact:faster_than_quasirandom}
 The hybrid model is always at least as fast as the quasi-random model implemented with identical lists.
\end{fact}

This simple, but useful observation follows from the fact that in the hybrid model every node acts as in the quasi-random model until it encounters an informed node. In this case, since we assumed all lists to be the same, the node becomes useless in the quasi-random model as all successive nodes on its list will have also been informed once it tries to call them. In the hybrid model, however, the node can still potentially inform uninformed nodes.

\begin{fact}
\label{fact:delaying}
 If a node is \emph{delayed}, i.e., halted for a number of rounds, then the protocol can only become slower.  
\end{fact}

It follows that, in our upper bound analysis, we can assume that a node is delayed any number of rounds. 



%

\begin{proof}[Proof of Theorem \ref{thm:tradeoff}]   
 We first analyze the time until all nodes are informed. 
We distinguish three phases of the rumor spreading process. The first phase lasts for $\log_2 n + h(n)$ rounds where $h(n)$ is an arbitrarily slowly growing function. By Fact \ref{fact:delaying}, we can assume that every node is delayed to the second phase once it contacts an informed node. Since this delayed protocol remains at least as fast as the hybrid model with $R = 1$, it follows from Fact \ref{fact:faster_than_quasirandom} that it is still faster than the quasi-random model implemented with identical lists. \citet{FH09} showed that the quasi-random model informs $(1-\e)n$ nodes for an arbitrarily small constant $\e > 0$ with probability $1-o(1)$ in this phase. Thus, we get the same result for our delayed protocol. 

The second phase lasts for $R$ rounds. By our delaying assumption, every node that is informed in the first phase will remain active for at least $R-1$ rounds before the second phase ends. The crucial observation is that every informed node that is still active either informs an uninformed node in a single round or calls a random node
in the next round. The former happens at most $\e n$ times in total.
We conclude that at the end of the second phase the nodes will have contributed at least $(1-\e)nR - \e n \geq (1-2\e)nR$ random calls (including the random calls made in the first phase). We show that then the largest interval of uninformed nodes is at most $(1+3\e)\ln (n)/R$. 
Let $I$ be an interval of length $(1+3\e)\ln (n)/R$. Then, the probability that no node in $I$ becomes informed in the second phase by these random calls is at most
\begin{align*}
 \biggl(1-\frac{(1+3\e)\ln n}{nR}\biggr)^{(1-2\e)nR} &\leq
\exp\left(-(1-2\e) (1+3\e) \ln n\right) \\							
&= n^{-1-\e+6\e^2} = n^{-1-\e'},
\end{align*}
for some constant $\e' > 0$ (when $\e$ is sufficiently small). Hence, by a union bound argument, it follows that there is no completely uninformed interval of
length $(1+3\e) \ln (n)/R$ after the second phase with probability at least $1-n^{-\e'}$ for some constant $\e' > 0$. 

In the last phase, all the remaining uninformed intervals are filled up. This takes at most the length of the largest uninformed interval which is
$(1+3\e) \ln (n)/R$ by the previous argument.

Using a simple union bound, we can bound the total failure probability by $o(1)$.   

It remains to bound the number of calls. Note that each node calls at most $R$ informed nodes in total. Hence, we use at most $n$ calls to inform all nodes and, in addition, at most $nR$ calls until all nodes stop informing.

\end{proof}

\subsubsection{Lower Bound}

In this section, we show that the upper bound from the previous section is essentially sharp. 

\begin{theorem}
 \label{thm:lowerbound}
Let $\e > 0$. If the hybrid model with $R$ random calls per node is run for less than 
\begin{align*}
 \log_2 (n) + (1-\e)\ln (n)/R +\tfrac{1}{2} R, &&\text{if $R \leq \sqrt{2(1-\e)\ln n}$,}\\
  \log_2 (n) + \sqrt{2(1-\e)\ln n}, &&\text{if $R \geq \sqrt{2(1-\e)\ln n}$}
\end{align*}
rounds, then with probability $1-\exp(-n^{\Theta(\eps)})$ not all nodes are informed.
\end{theorem}

\begin{proof}
Let $\Delta = \min\{(1-\e)\ln (n)/R+ \frac{1}{2}R, \sqrt{2(1-\e)\ln n}\}$ and $T = \log_2 n + \Delta$. 
We first analyze the probability that one (specific) vertex $u$ with a distance of more than $T$ from the starting node (in the cyclic order) becomes informed in the first $T$ rounds. We
call the event that a node chooses another node to inform uniformly at random a \emph{random call}. Hence, random calls occur as first calls of a node after the node encountered an already informed node. Clearly, $u$ remains
uninformed if for all $i \le T$ all random calls happening at time $T-i$ avoid $u$ and the $i$ vertices to the left of it. We say that $u$ is
\emph{unaffected} by such a random call.

We show that for $\e' > 0$
\begin{enumerate}
	\item[(i)] with probability at least $1 - 3 n^{-\eps'} \log_2 n$, $u$ is unaffected by random calls happening in
rounds $1$ to
$(1-\e')\log_2 n$,
	\item[(ii)] with probability at least $(1-\tfrac{\e' \log_2 (n) + \Delta}{n})^{n}$, $u$ is unaffected by random
calls happening in
rounds $(1-\e')\log_2 n +1$ to $\log_2 n$,
	\item[(iii)] with probability at least $\prod_{j=1}^{\min\{R,\Delta\}} \Bigl(1-\frac{\Delta-j}{n}\Bigr)^n$, $u$ is unaffected by random calls happening in
the remaining rounds $\log_2 (n)+1$ to $\log_2 (n) + \Delta$.
\end{enumerate}

We start with analyzing the effect of random calls happening in a first phase lasting for $(1-\e')\log_2 n$ rounds.  
In this phase, at most $n^{1-\e'}$ nodes can become informed since the number of informed nodes can at most
double in each round. Thus, we have at most $n^{1-\e'}$ random calls in this phase. The probability that a particular of
these calls
affects $u$ is at most $T/n$. 
Using a simple union bound, we conclude that the probability that any random call affects $u$ is at most $n^{1-\eps'} T/n
\le 3 n^{-\eps'} \log_2 n$.

For random calls happening in the second phase consisting of rounds $(1-\eps')\log_2 n + 1$ to $\log_2 n$, we argue as
follows. We bound from above the number of random calls by $n$ and the probability of each one affecting $u$ by $(\e' \log_2 (n) + \Delta)/n$. Since these are too many random decisions for a simple union bound, we use the fact that each random call addresses a node
chosen independently from previous random decisions. This yields $$\Pr(u \text{ is unaffected in second phase}) \geq \Bigl(1-\tfrac{\e' \log_2 (n) + \Delta}{n}\Bigr)^{n}.$$

Note that in the third phase we have at most $\min\{R,\Delta\}$ random calls per node. The probability that the $i$-th random call of a node in round $\log_2 (n)+1$ to $\log_2 (n) + \Delta$ affects $u$ is at most $(\Delta-i)/n$. Hence, the probability that $u$ is unaffected in the third phase is at least $\Bigl(\prod_{j=1}^{\min\{R,\Delta\}} (1-\frac{\Delta-j}{n})\Bigr)^n$. We distinguish two cases:
If $\min\{R,\Delta\} = R$, then 
\begin{align}
\Pr(u \text{ is unaffected in third phase}) &\geq \prod_{j=1}^{R} \Bigl(1-\frac{\Delta-j}{n}\Bigr)^n \notag\\
			      &\geq \exp(\sum_{j=1}^{R} -\Delta+j- (\Delta - j)^2/n) \label{i1}\\ 
			     &\geq (1-o(1))\exp(\sum_{j=1}^{R} -\Delta+j) \label{i2}\\
			   &\geq (1-o(1))\exp( -R\Delta+ R(R+1)/2) \notag
			  \geq (1-o(1))n^{-1+\e}\notag,
\end{align}
where \eqref{i1} follows from the fact that for $x \leq \tfrac{1}{2}$, we have  $1-x \geq e^{-x-x^2}$, \eqref{i2} follows from our assumption $R \leq \Delta \leq O(\sqrt{\ln n})$, and the last inequality follows from the definition of $\Delta$.
Similarly, if $\min\{R,\Delta\} = \Delta$, then
\begin{align*}
\Pr(u \text{ is unaffected in third phase}) &\geq \prod_{j=1}^{\Delta} \Bigl(1-\frac{\Delta-j}{n}\Bigr)^n
			   \geq (1-o(1))\exp( -\Delta^2/2) \geq (1-o(1))n^{-(1-\e)}.
\end{align*}
We set $\e' := \e/4$. Since all random calls choose their addressees independently, we have
\begin{align}
  \Pr(u &\text{ remains unaffected in all three phases}) \notag \\
  &\ge  (1-3 n^{-\eps'} \log_2 (n)) \Bigl(1-\tfrac{\e' \log_2 n + \Delta}{n}\Bigr)^{n} \prod_{j=1}^{\min\{R,\Delta\}} \Bigl(1-\frac{\Delta-j}{n}\Bigr)^n \notag \\
  &\ge  (1-o(1)) \exp(-2\eps' \log_2 n - (1-\eps)\ln n)
  \ge  n^{-1+\Theta(\eps)} \notag.
\end{align}

Let $k = \frac{n}{T+1} -1$. Let $u_1, \dots, u_k$ be nodes each having distance more than $T$ from each other and from the starting node (in the cyclic order). We argue that, with sufficiently high probability, one such node will remain uninformed after the first $T$ rounds. Let $U_i$ denote the event that node $u_i$ is informed. Note that since these nodes have a
distance of $T$ from each other, a random call that informs one such node $u_i$ can not lead to the informing of any other node $u_j$ during the first $T$ rounds. Hence, these events are \emph{negatively correlated}: if some nodes are informed, the probability that another one is also informed decreases, or formally, $\Pr(U \mid U_1, \dots, U_j) \leq \Pr(U)$. 
We compute 
\begin{align*}
\Pr(\text{no node remains uninformed})  \leq \Pr(U_1 \wedge \dots \wedge U_{k}) \leq \prod_{1 \leq j\leq k} \Pr(U_j) 
					\leq (1- n^{-1+\Theta(\eps)})^{k}
				      \leq \exp({-n^{\Theta(\eps)}}).\end{align*}
\end{proof}

\bibliographystyle{myabbrvnat}
\bibliography{broadcast}

\end{document}